\documentclass[12pt]{article}
\begin{document}
\title{Runge-Lenz operator for Dirac field in 
Taub-NUT background}

\author{Ion I. Cot\u aescu \thanks{E-mail:~~~cota@quasar.physics.uvt.ro}\\ 
{\small \it West University of Timi\c soara,}\\
       {\small \it V. P\^ arvan Ave. 4, RO-1900 Timi\c soara, Romania}
\and
Mihai Visinescu \thanks{E-mail:~~~mvisin@theor1.theory.nipne.ro}\\
{\small \it Department of Theoretical Physics,}\\
{\small \it National Institute for Physics and Nuclear Engineering,}\\
{\small \it P.O.Box M.G.-6, Magurele, Bucharest, Romania}}
\date{\today}

\maketitle

\begin{abstract}
Fermions in $D=4$ self-dual Euclidean Taub-NUT space are investigated. 
Dirac-type operators involving Killing-Yano tensors of the Taub-NUT 
geometry are explicitly given showing that they anticommute with the 
standard Dirac operator and commute with the Hamiltonian  as it is 
expected. They are connected with the hidden symmetries of the space 
allowing the construction of a conserved vector operator analogous to 
the Runge-Lenz vector of the Kepler problem. This operator is written 
down pointing out its algebraic properties.  

Pacs 04.62.+v

\end{abstract}

\section{Introduction}

It is highly non-trivial problem to study the Dirac equation in a curved 
background. One of the most interesting  geometries is that of the Euclidean 
Taub-NUT space  since it has not only usual isometries but also it admits a 
hidden symmetry of the Kepler type  \cite{GM,GRFH}. Moreover in the Taub-NUT 
geometry there are four Killing-Yano tensors, connected with three 
St\" ackel-Killing tensors  which represent the components of an analogue to 
the Runge-Lenz vector of the Kepler problem \cite{GRFH,vH1,VV,MV}.

In other respects, the Taub-NUT metric is involved in many modern studies in 
physics \cite{GM,GRFH,MAH}. The Dirac equation in the Kaluza-Klein monopole 
field was studied in the mid eighties \cite{DIRAC} but, to our knowledge,  
the conserved observables of the Dirac theory corresponding to the hidden 
symmetries of the Taub-NUT geometry have not been well discussed in  
literature. An attempt to take into account the Runge-Lenz vector of this 
geometry was done in \cite{CH}, but their approach is incomplete. 

For this reason the present paper is devoted to the construction of the 
Runge-Lenz vector-operator for the  Dirac field in the 
Taub-NUT background. The method we use is to calculate the Dirac-type 
operators defined with the help of the  Killing-Yano tensors and then to 
write down the Runge-Lenz operator.    

It is known that space-time isometries give rise to constants of motion 
along geodesics. However not all conserved quantities along geodesics 
arise from isometries. There are prime integrals of motion  related to 
{\em hidden} symmetries of the manifold, a manifestation of the existence 
of the St\" ackel-Killing tensors \cite{Ca1,CRGH}.
A St\" ackel-Killing tensor of valence $r$ is a tensor $k_{\mu_1...\mu_r}$  
which is completely symmetric and which satisfies a generalized Killing 
equation $k_{(\mu_1...\mu_r;\lambda)} = 0$. 

There are geometries where the St\" ackel-Killing tensor can have a certain 
root represented by Killing-Yano tensors. We recall that a tensor 
$f_{\mu_1...\mu_r}$ is a Killing-Yano tensor of valence $r$  if it is totally 
antisymmetric and  satisfies the equation 
$f_{\mu_1...(\mu_r;\lambda)} = 0$ \cite{Ya}.
The role of the Killing-Yano tensors becomes important in theories with spin
as in the cases of the pseudo-classical  spinning manifolds \cite{BM,GRH} 
or in quantum theory of fermions \cite{CV2}.

On the other hand, Carter and McLenaghan \cite{CML} showed that, in the theory 
of Dirac fermions, for any isometry with Killing vector $k^\mu$ there is an 
appropriate operator 
\begin{equation}\label{kil}
X_k = -i ( k^\mu \hat\nabla_\mu - {1\over 4} \gamma^\mu \gamma^\nu 
k_{\mu;\nu})
\end{equation}
which {\em commutes}  with the  {\em standard} Dirac operator 
defined by
\begin{equation}\label{ds}
D_s = \gamma^\rho \hat\nabla_\rho
\end{equation}
where $\hat\nabla_\rho$ are the spin covariant derivatives including the  
spin-con\-nec\-ti\-on while $\gamma^\mu$ are the point-dependent  Dirac 
matrices carrying natural indices. The usual Dirac matrices with local 
(hated) indices will be denoted by $\hat\gamma^{\hat\mu}$. We specify that
$X_{k}$ is just the generator of the 
operator-valued representation according which the Dirac field must transform 
under the isometry transformations corresponding to $k^{\mu}$ \cite{ES}. 
Moreover, as stated in \cite{CML}, each Killing-Yano tensor $f_{\mu\nu}$ 
of valence 2 produces the non-standard Dirac operators of the form 
\begin{equation}\label{df}
D_f = -i\gamma^\mu (f_\mu ^{~\nu}\hat\nabla_\nu  - 
{1\over 6}\gamma^\nu 
\gamma^\rho f_{\mu\nu;\rho})
\end{equation}
which {\em anticommutes} with the standard Dirac operator $D_s$. We observe 
that the operators (\ref{kil}) and (\ref{df}) have specific spin parts. 

\section{The Dirac field in Taub-NUT space}

Let us consider the Taub-NUT space and the chart with Cartesian coordinates 
$x^{\mu}$  ($\mu, \nu,...=1,2,3,4$) having the line element  
\begin{equation}\label{(met)} 
ds^{2}=g_{\mu\nu}dx^{\mu}dx^{\nu}=\frac{1}{V}dl^{2}+V(dx^{4}+
A_{i}dx^{i})^{2}
\end{equation}   
where $dl^{2}=(d\vec{x})^{2}=(dx^{1})^{2}+(dx^{2})^{2}+(dx^{3})^{2}$
is the  Euclidean three-dimen\-sio\-nal line element and $\vec{A}$ is 
the gauge field of a monopole. Another chart 
suitable for applications is that of spherical coordinates, $x'=
(r,\,\theta,\,\phi,\,\chi)$, among them the first three are the  
spherical coordinates commonly associated with the  Cartesian space 
ones, $x^{i}$  ($i,j,...=1,2,3$), while 
$\chi+\phi=-\mu x^{4}$. The real number $\mu$ is the  parameter of the 
theory which enters in the form of the function $1/V(r)=1+\mu/r$. 
The unique non-vanishing component of the vector potential in 
spherical coordinates is $A_{\phi}=\mu(1-\cos\theta)$. In Cartesian 
coordinates we get  ${\rm div}\vec{A}=0$ and $\vec{B}\,={\rm rot}\, 
\vec{A}=\mu\frac{\vec{x}}{r^3}$.

When one uses Cartesian charts in the Taub-NUT geometry, it is useful 
to consider the local frames given by tetrad fields, $e(x)$ and $\hat e(x)$, 
as defined in \cite{P}. Their components, have the usual orthonormalization 
properties and give the components of the metric tensor,  
$g_{\mu\nu}=\eta_{\hat\alpha \hat\beta}\hat 
e^{\hat\alpha}_{\mu}\hat e^{\hat\beta}_{\nu}$ and 
$g^{\mu\nu}=\eta^{\hat\alpha \hat\beta} e_{\hat\alpha}^{\mu}
e_{\hat\beta}^{\nu}$ where 
$\eta_{\hat\alpha \hat\beta} =\delta_{\hat\alpha \hat\beta} $ is the 
Euclidean  metric tensor.  Its gauge group,  $G(\eta)=SO(4)$,  
has  the  universal covering group     
$\tilde G(\eta)\sim SU(2)\otimes SU(2)$. The four Dirac matrices 
matrices  $\hat\gamma^{\hat\alpha}$, that  satisfy 
$\{ \hat\gamma^{\hat\alpha},\, \hat\gamma^{\hat\beta} \}
=2\eta^{\hat\alpha \hat\beta}$, can be taken as  
\begin{equation}\label{(gammai)} 
\hat\gamma^i = -i
\left(
\begin{array}{cc}
0&\sigma_i\\
-\sigma_i&0
\end{array}\right)\,,\quad  
\hat\gamma^4 =
\left(
\begin{array}{cc}
0&{\bf 1}_2\\
{\bf 1}_2&0
\end{array}\right)\,.
\end{equation}
In addition we consider the matrix
\begin{equation}\label{(gamma5)} 
\hat\gamma^5 = \hat\gamma^1\hat\gamma^2\hat\gamma^3\hat\gamma^4 =
\left(
\begin{array}{cc}
{\bf 1}_2&0\\
0&-{\bf 1}_2
\end{array}\right)
\end{equation}
which is denoted by $\gamma^{0}$ in Kaluza-Klein theory explicitly involving 
the time \cite{CV2}.
These matrices are self-adjoint  
and give the covariant basis generators of the group $\tilde G(\eta)$, 
denoted by  $S^{\hat\alpha \hat\beta}=-i 
[\hat\gamma^{\hat\alpha},\, \hat\gamma^{\hat\beta}]/4$.

The standard Dirac operator is defined as \cite{CV2}
\begin{equation}\label{(de)}
{D}_{s}=\hat\gamma^{\hat\alpha}\hat\nabla_{\hat\alpha}  
=i\sqrt{V}\vec{\hat\gamma}\cdot\vec{P}
+\frac{i}{\sqrt{V}}\hat\gamma^{4}P_{4}
+\frac{i}{2} V\sqrt{V}\hat\gamma^{4}\vec{\Sigma}^{*}\cdot\vec{B}
\end{equation}
where  $\hat\nabla_{\hat\alpha}$ are the components of the  
spin covariant derivatives with local indices 
\begin{equation}\label{sder}
\hat\nabla_{i}=i\sqrt{V}P_{i}+\frac{i}{2}V\sqrt{V}\varepsilon_{ijk}
\Sigma_{j}^{*}B_{k}\,,\quad
\hat\nabla_{4}=\frac{i}{\sqrt{V}}P_{4}-\frac{i}{2}V\sqrt{V}
\vec{\Sigma}^{*}\cdot\vec{B}\,.
\end{equation}
These depend on the momentum operators  
$P_{i}=-i(\partial_{i}-A_{i}\partial_{4})$ and $P_{4}=-i\partial_{4}$\,,
which obey the commutation rules
$[P_{i},P_{j}]=i\varepsilon_{ijk}B_{k}P_{4}$ and
$[P_{i},P_{4}]=0$\,. 
The spin matrices giving the spin connection are  
\begin{equation}\label{sigst}
\Sigma_{i}^{*}=S_{i}+\frac{i}{2}\hat\gamma^{4}\hat\gamma^{i} \,, \quad
S_{i}=\frac{1}{2}\varepsilon_{ijk}S^{jk}\,.
\end{equation}

In our representation of the Dirac matrices, the Hamiltonian operator of the 
{\em massless} Dirac field reads \cite{CV2}
\begin{equation}\label{HH}
H =\hat\gamma^5{D}_{s}=\left(
\begin{array}{cc}
0&V\pi^{*}\frac{\textstyle 1}{\textstyle \sqrt{V}}\\
\sqrt{V}\pi&0
\end{array}\right)\,.
\end{equation}
This is expressed in terms of the operators 
$\pi=\,{\sigma}_{P}-iV^{-1}P_{4}$ and 
$\pi^{*}=\,{\sigma}_{P}+iV^{-1}P_{4}$ 
where $\sigma_P=\vec{\sigma}\cdot\vec{P}$ 
involves the Pauli matrices $\sigma_i$. Denoting by 
$\nabla_{\mu}$ the usual covariant derivatives we find that the Klein-Gordon 
operator has the equivalent forms   
\begin{equation}
\Delta= -\nabla_{\mu}g^{\mu\nu}\nabla_{\nu}=V\,\pi^{*}\pi= 
V{\vec{P}\,}^{2}+\frac{1}{V}{P_{4}}^{2}\,.
\end{equation}

\section{Conserved observables}

We say that an operator represents a {\em conserved} observable if 
this {\em commutes} with $H$. 
The specific off-diagonal form of $H$ suggested us to introduce the 
${\cal Q}$-operators defined as \cite{CV2}
\begin{equation}
{\cal Q}(X)=\left\{ H\,,\,\left(
\begin{array}{cc}
X&0\\
0&0
\end{array} 
\right) \right\}
=i\left[ Q_{0}\,,\,\left(
\begin{array}{cc}
X&0\\
0&0
\end{array} 
\right) \right] 
\end{equation}
where $Q_0=i{D}_s=i\hat\gamma^{5}H$. These operators  
allow us to associate Dirac operators to the Pauli operators, 
$\pi,\,\pi^{*}$, $\sigma_{P}$, \,$\sigma_{L}=\,\vec{\sigma}\cdot\vec{L}$ and 
$\sigma_{r}=\vec{\sigma}\cdot\vec{x}/r$ \cite{DYON}.
The remarkable property of the ${\cal Q}$-operators is that if 
$[X,\,\Delta]=0$ then ${\cal Q}(X)$ commutes with $H$ \cite{CV2}. 
Other conserved observables can be of diagonal form  
$T={\rm diag}\,(T^{(1)},\,T^{(2)})$  when the condition 
$[T,\,H]=0$ implies   
\begin{equation}\label{T12}
T^{(2)}\sqrt{V}\pi=\sqrt{V} \pi T^{(1)}
\,,\quad V\pi^{*}\frac{1}{\sqrt{V}}T^{(2)}=
T^{(1)}V\pi^{*}\frac{1}{\sqrt{V}}
\end{equation}
and, therefore,  $[T^{(1)},\,\Delta]=0$.

The conserved observables may be found among the operators 
which commute or anticommute  simultaneously with $D_s$  and $\hat\gamma^5$. 
An important type of such operators are the generators of the 
operator-valued representations of the isometry group carried by the spaces 
of physical fields \cite{CML,ES}. The space defined by (\ref{(met)}) has  
the isometry group $G_{s}=SO(3)\otimes U_{4}(1)$, 
of rotations of the coordinates $x^i$  and translations of $x^{4}$,
with the Killing vectors $k_i$ ($i=1,2,3$) and $k_4$ respectively.
According to Eq.(\ref{kil}) the $U_{4}(1)$ generator is $P_{4}$  
while other three Killing vectors  give the 
$SO(3)$ generators which are the components of the whole angular momentum 
$\vec{J} =\vec{L} + \vec{S}$  as in the flat space-times. The difference 
is that here the orbital angular momentum is  
\begin{equation}\label{(angmom)}
\vec{L}\,=\,\vec{x}\times\vec{P}-\mu\frac{\vec{x}}{r}P_{4}\,.
\end{equation} 
The components of $\vec{J}$ commute with ${D}_s$ and 
$\hat\gamma^5$ and, therefore, they are conserved, commuting with $H$.   
Moreover, they  satisfy the canonical commutation rules 
among themselves and with the components of all the other vector operators 
(e.g. coordinates,  momenta, spin etc.). 

The first three Killing-Yano tensors of the Taub-NUT space \cite{GRFH}
\begin{equation}\label{fi}
f^i 
= f^i_{\,{\hat \alpha}{\hat \beta}} {\hat e}^{\hat \alpha} \wedge 
{\hat e}^{\hat \beta}
= 2 {\hat e}^5\wedge  {\hat e}^i +\varepsilon_{ijk} {\hat e}^j\wedge 
{\hat e}^k
\end{equation}
are rather special since they are covariantly constant. According to 
Eq.(\ref{df}), after some algebra, we obtain the Dirac-type operators  
\begin{equation}
Q_{i}=-if^{i}_{\,\hat\alpha\hat\beta}\hat\gamma^{\hat\alpha}\hat\nabla
^{\hat\beta}={\cal Q}(\sigma_{i})
\end{equation}
which anticommute with $Q_{0}$ and $\hat\gamma^5$, commute with $H$ and 
obey the $N=4$ superalgebra 
\begin{equation}\label{QQH}
\{Q_{A},\,Q_{B}\}=2\delta_{AB}H^2\,, \quad A,B,...=0,1,2,3
\end{equation}  
linked to the hyper-K\" ahler geometry of the Taub-NUT space.

The fourth Killing-Yano tensor of the Taub-NUT space
\begin{equation}\label{fY}
f^{Y}=-\frac{x^i}{r}f^{i}+\frac{2x^i}{\mu V}\varepsilon_{ijk}\hat e^{j}\land
\hat e^{k}
\end{equation}
is not covariantly constant having the following non-vanishing field 
strength components
\begin{equation}\label{stren}
f^{Y}_{\,r\theta;\phi}=\frac{2r^2}{\mu V}\sin\theta.
\end{equation}
Taking into account these covariant derivatives we can calculate its 
corresponding Dirac-type operator, $Q^Y$, according to the 
general rule (\ref{df}), as indicated in  Appendix. The  result is
\begin{equation}\label{dy2}
Q^Y=\frac{r}{\mu}\left\{H\,,\left(
\begin{array}{cc}
\sigma_{r}&0\\
0&-\sigma_{r} V^{-1}
\end{array}\right)
\right\}
=i\frac{r}{\mu}\left[Q_{0}\,,\left(
\begin{array}{cc}
\sigma_{r}&0\\
0& \sigma_{r} V^{-1}
\end{array}\right)
\right]\,.
\end{equation}
One can verify that this operator commutes with $H$ and anticommutes with 
$Q_{0}$ and $\hat\gamma^5$.

\section{The Runge-Lenz operator} 

The hidden symmetries of the Taub-NUT geometry are encapsulated in 
the non-trivial St\" ackel-Killing tensors $k_{i\mu\nu}, (i=1,2,3)$.
They can be expressed as symmetrized products of Killing-Yano tensors 
(\ref{fi}) and  (\ref{fY}) \cite{VV}:
\begin{equation}\label{kff}
k_{i\mu\nu} = -{\mu\over 4}(f^Y_{~\mu\lambda}f^{i\lambda}_{~~\nu} +
f^Y_{~\nu\lambda}f^{i\lambda}_{~~\mu})+{1\over 2\mu}(k_{4\mu}k_{i\nu} +
k_{4\nu}k_{i\mu}).
\end{equation}
In fact only the product of Killing-Yano tensors $f^i$ and $f^Y$ leads 
to non-trivial St\" ackel-Killing tensors, the last term in the r.h.s. 
of (\ref{kff}) being a simple product of Killing vectors. This term is 
usually added to write the Runge-Lenz vector of the scalar (Schr\" odinger 
or Klein-Gordon) theory \cite{GRFH,CV1}
\begin{equation}\label{RLorb}   
\vec{K}=-\frac{1}{2}\nabla_{\mu}\vec{k}^{\mu\nu}\nabla_{\nu}=
\frac{1}{2}(\vec{P}\times \vec{L}-\vec{L}\times \vec{P})-
\frac{\mu}{2}\frac{\vec{x}}{r}\Delta +\mu\frac{\vec{x}}{r}{P_{4}}^{2}
\end{equation}
whose components commute with $\Delta$ and obey 
\begin{equation}\label{comk}
\left[ L_{i},\, K_{j} \right] = i \varepsilon_{ijk}\,K_{k}~~~,~~~ 
\left[ K_{i},\, K_{j} \right] = i ({P_4}^2-\Delta)\varepsilon_{ijk}L_{k}\,. 
\end{equation}

For the Dirac theory the construction of the Runge-Lenz operators can 
be done using products among the operators $Q^Y$ and $Q_i$. This 
procedure represents the only possibility to generate non-trivial 
conserved operators which should be not related to the Hamiltonian, like in 
Eq.(\ref{QQH}), or the Casimir operators of the symmetry group of the 
manifold.

Using an analogy with the relation (\ref{kff}) from the Taub-NUT 
geometry, let us consider the vector operator $\vec{N}$ with the  
components  
\begin{equation}
{N}_{i}=\frac{\mu}{4}\left\{ Q^{Y},\, Q_{i}\right\}-J_{i}P_{4}
\end{equation}
which, after simple but tedious calculations based on Eqs. (\ref{A1}) and 
(\ref{A2}) from Appendix, can be put in the form
\begin{eqnarray}\label{N}
\vec{N}&=&\frac{1}{2}W\left(\vec{P}\times \vec{J}-\vec{J}\times \vec{P}-
i\hat\gamma^{4}\vec{\hat\gamma}\times \vec{P}\right)W^{-1}-
\frac{\mu}{2}{\cal Q}(\vec{x}/r)H\nonumber\\ 
&&+\left[\vec{S}+\vec{x}\times(\vec{\Sigma}^{*}\times \vec{B})\right]P_{4}
+\mu\frac{\vec{x}}{r}{P_{4}}^{2} \label{NN}
\end{eqnarray}
where $W={\rm diag}\,(1,\,\sqrt{V})$ and $\vec{\Sigma}^{*}$ is given by 
(\ref{sigst}).

Since $\vec{N}$ commutes with $H$, its diagonal  blocks,   
$\vec{N}^{(1)}$ and $\vec{N}^{(2)}$, satisfy the relations
(\ref{T12}). It is straightforward to find that the first block which 
commutes with $\Delta$,
\begin{equation}
\vec{N}^{(1)}=\vec{K}+\frac{\vec{\sigma}}{2}P_{4}\,,
\end{equation}
contains not only the orbital Runge-Lenz operator (\ref{RLorb}) but a 
spin term too. Moreover the components of the operator  $\vec{N}$
satisfy the following commutation relations:
\begin{eqnarray}
\left[{N}_{i},\,P_{4}\right]=0\,,&\quad&
\left[{N}_{i},\,J_{j}\right]=i\varepsilon_{ijk}{N}_{k}\,,\\
\left[{N}_{i},\,Q_{0}\right]=0\,,&\quad&
\left[{N}_{i},\,Q_{j}\right]=i\varepsilon_{ijk}Q_{k}P_4 \label{nq}
\end{eqnarray}
and
\begin{equation}
\left[{N}_{i},\,{N}_{j}\right]= i\varepsilon_{ijk}J_{k}F^2
+\frac{i}{2}\varepsilon_{ijk} Q_{i}H\,, \quad F^2={P_4}^2-H^2\,.
\end{equation}

In order to put the last commutator in a form close to (\ref{comk})
we can redefine the components of the Runge-Lenz operator, $\vec{\cal K}$, 
as follows
\begin{equation}\label{RL}
{\cal K}_{i}={N}_{i}+ \frac{1}{2}H^{-1}(F-P_4) Q_i
\end{equation}
These operators satisfy the desired commutation rules 
\begin{eqnarray}
\left[{\cal K}_{i},\,H\right]=0\,,&\quad&
\left[{\cal K}_{i},\,J_{j}\right]=i\varepsilon_{ijk}{\cal K}_{k}\,,\\
\left[{\cal K}_{i},\,P_4\right]=0\,,&\quad&
\left[{\cal K}_{i},\,Q_{j}\right]=i\varepsilon_{ijk}Q_{k}F
\end{eqnarray}
and
\begin{equation}
\left[{\cal K}_{i},\,{\cal K}_{j}\right]= i\varepsilon_{ijk}J_{k}F^2\,.
\end{equation}

The explicit form of them makes obvious the spin contribution to the 
Runge-Lenz vector, re-confirming the result from pseudo-classical approach 
\cite{VV,MV1}.
We specify that there are no zero modes \cite{CV2} and, therefore, the 
operator $H$ is invertible such that our definition (\ref{RL}) of the 
Runge-Lenz operator is correct. Moreover, as in the scalar case 
\cite{MAH,GRFH}, this can be re-scaled in order to recover the familiar 
dynamical algebras $o(4)$, $o(3,1)$ or $e(3)$, corresponding to different 
spectral domains of the Kepler-type problems.  

Finally, it is worthy to note that commutation relation (\ref{nq}) of 
the Runge-Lenz operator with the standard Dirac operator remains valid 
even if a mass term is included in Eq.(\ref{ds}).

\setcounter{equation}{0} \renewcommand{\theequation} 
{A.\arabic{equation}}

\section*{Appendix: The operator $Q^Y$}

To evaluate the operator $Q^Y$ we start with the first term of (\ref{df})
\begin{equation}\label{1t}
-if^Y_{\,\hat\alpha\hat\beta}\hat\gamma^{\hat\alpha}\hat\nabla
^{\hat\beta}=-\frac{x^i}{r} Q_{i}+\frac{2i}{\mu\sqrt{V}}
\left(
\begin{array}{cc}
0&\lambda-V\\
1-\lambda&0
\end{array}
\right)\,.
\end{equation}
where $\lambda =\vec{\sigma}\cdot(\vec{x}\times\vec{P})+1=\sigma_{L}+1+
\mu\sigma_{r}P_{4}$ is the operator introduced in \cite{JMP}.
Since the Dirac matrices with spherical indices,   
$\gamma^{\mu}(x')={e'}^{\mu}_{\hat\alpha}(x')\hat\gamma^{\hat\alpha}$, 
corresponding to the orthogonal coordinates $r,\,\theta$ and $\phi$ are
\begin{eqnarray}
\gamma^{r}(x')&=&\sqrt{V}(\hat\gamma^{1}\sin\theta\cos\phi+
\hat\gamma^{2}\sin\theta\sin\phi +\hat\gamma^{3}\cos\theta)\,,\\            
\gamma^{\theta}(x')&=&r^{-1}\sqrt{V}(\hat\gamma^{1}\cos\theta\cos\phi+
\hat\gamma^{2}\cos\theta\sin\phi -\hat\gamma^{3}\sin\theta)\,,\\            
\gamma^{\phi}(x')&=&(r\sin\theta)^{-1}\sqrt{V}(-\hat\gamma^{1}\sin\phi+
\hat\gamma^{2}\cos\phi)
\end{eqnarray}
we find that 
\begin{equation}\label{3gama}
\gamma^{r}\gamma^{\theta}\gamma^{\phi}=
-\gamma^{r}\gamma^{\phi}\gamma^{\theta}=\cdots=
r^{-2}(\sin\theta)^{-1}V\sqrt{V}\,\hat\gamma^{5}\hat\gamma^{4}
\end{equation}
is completely antisymmetric in $r,\,\theta,\,\phi$. Then, we calculate the 
second term of (\ref{df}), according to (\ref{stren}) and (\ref{3gama}), 
and embedding the result with (\ref{1t}) we get
\begin{equation}\label{dy1}
Q^Y=-{\cal Q}(\sigma_r)+\frac{2i}{\mu\sqrt{V}}
\left(
\begin{array}{cc}
0&\lambda\\
-\lambda&0
\end{array}
\right)\,.
\end{equation}
The form (\ref{dy2}) and other interesting formulas can be derived  using the 
identities $[\sigma_r,\, \sigma_P]=2ir^{-1}\lambda$ and
\begin{equation}
\sigma_P\lambda=-\lambda\sigma_P=\frac{i}{2}\,\vec{\sigma}\cdot(\vec{P}\times
\vec{L}-\vec{L}\times\vec{P})-\frac{i\mu}{r}\lambda P_{4}\,.
\end{equation}
With their help one can demonstrate that the relations
\begin{eqnarray}
Q^Y{\cal Q}(X)&=&-{\cal Q}(\sigma_{r}X)H+\frac{2i}{\mu}\left(
\begin{array}{cc}
\lambda\pi X&0\\
0&-\sqrt{V}\lambda X\pi^{*}\frac{\textstyle 1}{\textstyle \sqrt{V}}\\
\end{array}\right)\,,\label{A1}\\
{\cal Q}(X)Q^{Y}&=&-{\cal Q}(X\sigma_{r})H+\frac{2i}{\mu}\left(
\begin{array}{cc}
-X\pi^{*}\lambda&0\\
0&\sqrt{V}\pi X\lambda\frac{\textstyle 1}{\textstyle \sqrt{V}}\\
\end{array}\right) \label{A2}
\end{eqnarray}
hold for any $2\times 2$ matrix operator $X$ which commutes with $\Delta$ and 
$V$. In particular for $X=1$ it results $[Q^{Y},\,H]=0$.


\begin{thebibliography}{99}
%1
\bibitem{GM}
G. W. Gibbons and N. S. Manton,
{\em Nucl. Phys.} {\bf B274} (1986) 183.
%2
\bibitem{GRFH}
G. W. Gibbons and P. J. Ruback, 
{\em Phys. Lett.} {\bf B188} (1987)226;
{\em Commun. Math. Phys.} {\bf 115} (1988) 267;
L. Gy. Feher and P. A. Horv\' athy, 
{\em Phys. Lett.} {\bf B183} (1987) 182; 
B. Cordani, Gy. Feher and P. A. Horv\' athy,
{\em Phys. Lett.} {\bf B201} (1988) 481.
%3
\bibitem{vH1}
J. W. van Holten, {\em Phys. Lett.} {\bf B342} (1995) 47.
%4
\bibitem{VV}
D. Vaman and M. Visinescu, 
{\em Phys. Rev.} {\bf D57} (1998) 3790;
D. Vaman and M. Visinescu, 
{\em Fortschr. Phys.} {\bf 47} (1999) 493.
%5
\bibitem{MV}
M. Visinescu,
{\em J. Phys. A: Math. Gen.} {\bf 33} (2000) 4383.
%6
\bibitem{MAH}
N. S. Manton,
{\em Phys. Lett.} {\bf B110} (1985) 54; 
{\em Phys. Lett.} {\bf B154} (1985) 397; 
M. F. Atiyah and  N. Hitchin, {\em Phys. Lett. } {\bf A107} (1985) 21.
%7
\bibitem{DIRAC}
Z. F. Ezawa and A. Iwazaki, {\em Phys. Lett.} {\bf 138B} (1984) 81; 
M. Kobayashi and A. Sugamoto, {\em Progr. Theor. Phys.} {\bf 72} (1984) 
122; A. Bais and P. Batenberg, {\em Nucl. Phys.} {\bf B245} (1984) 469.
%8
\bibitem{CH}
A. Comtet and P. A. Horv\' athy, {\em Phys. Lett.} {\bf B349} (1995) 49.
%9
\bibitem{Ca1}
B. Carter, {\em Phys. Rev.} {\bf 16} (1977) 3395.
%10
\bibitem{CRGH}
M. Crampin, {\em Rep. Math. Phys.} {\bf 20} (1984) 31;
K. Rosquist, {\em J. Math. Phys.} {\bf 30} (1989) 2319;.
G. W. Gibbons and C. A. R. Herdeiro, {\em Class. Quant. Grav.}
{\bf 16} (1999) 3619.
%11
\bibitem{Ya}
K. Yano, 
{\em Ann. Math.} {\bf 55} (1952) 328.
%12
\bibitem{BM}
F. A. Berezin and M. S. Marinov, 
{\em Ann. Phys. N.Y.} {\bf 104} (1977) 336.
%13
\bibitem{GRH}
G. W. Gibbons, R. H. Rietdijk and J. W.van Holten, 
{\em Nucl. Phys.} {\bf B404} (1993) 42.
%14
\bibitem{CV2}
I. I. Cot\u aescu and M. Visinescu,
{\tt hep-th/0008181}; {\em Int. J. Mod. Phys. A} (in press).
%15
\bibitem{CML}
B. Carter and R. G. McLenaghan, {\em Phys. Rev.} {\bf D19} (1979) 1093.
%16
\bibitem{ES}
I. I. Cot\u aescu, {\em J. Phys. A: Math. Gen.} {\bf 33} (2000) 9177 
%17
\bibitem{P}
H. Boutaleb - Joutei and A. Chakrabarti, {\em Phys. Rev.} {\bf D21} (1979)
2280.
%18
\bibitem{DYON}
E. D'Hoker and L. Vinet, {\em Phys. Lett.} {\bf B137} (1984) 72;
F. Bloore and P. A. Horv\' athy, {\em J. Math. Phys.} {\bf 33} (1992) 
q1869.  
%19
\bibitem{CV1}
I. I. Cot\u aescu and M. Visinescu,
{\em Mod. Phys. Lett. A}{\bf 15} (2000) 145.
%20
\bibitem{MV1}
M. Visinescu, {\em Phys. Lett.} {\bf B339} (1994) 28.
%21
\bibitem{JMP}
F. De Jonge, A. J. Macfarlane, K. Peters and J.-W. wan Holten, 
{\em Phys. Lett.}  {\bf B359} (1995) 114.

\end{thebibliography}
\end{document}